# Laser control of an excited-state vibrational wave packet in neutral $H_2$

**Authors:**

Gergana D. Borisova[1*], Paula Barber Belda[1], Shuyuan Hu[1], Paul Birk[1], Veit Stooß[1], Maximilian Hartmann[1], Daniel Fan[1], Robert Moshammer[1], Alejandro Saenz[2], Christian Ott[1†] and Thomas Pfeifer[1#]

*borisova@mpi-hd.mpg.de

†christian.ott@mpi-hd.mpg.de

#thomas.pfeifer@mpi-hd.mpg.de

**Affiliations:**

[1]Max-Planck-Institut für Kernphysik, Saupfercheckweg 1, 69117 Heidelberg, Germany

[2]Institut für Physik, Humboldt-Universität zu Berlin, 12489 Berlin, Germany

**Abstract:**

**We observe and control a molecular vibrational wave packet in an electronically excited state of the neutral hydrogen molecule. In an extreme-ultraviolet (XUV) transient-absorption experiment we launch a vibrational wave packet in the $D\ {}^1\Pi_u\ 3p\pi$ state of $H_2$ and observe its time evolution via the coherent dipole response. The reconstructed time-dependent dipole from experimentally measured XUV absorption spectra provides access to the revival of the vibrational wave packet, which we control via an intense near-infrared (NIR) pulse. Tuning the intensity of the NIR pulse we observe the revival of the wave packet to be significantly modified, which is supported by the results of a multi-level simulation. The NIR field is applied only 7 fs after the creation of the wave packet but influences its evolution up to at least its first revival at 270 fs. This experimental approach for nonlocal-in-time laser control of quantum dynamics is generally applicable to a large range of molecules and materials as it only requires the observation of absorption spectra.**

**Main text:**

Ever since the emergence of ultrashort light pulses, pump-probe techniques provide information about the time-dependent dynamics in atomic and molecular systems [1–6]. Here, the first laser pulse (pump) initiates coherent dynamics by exciting a bound and/or a continuum wave packet (WP), and later the probe pulse is considered to take a "snapshot" of the system's time evolution at a specific time, yielding insight into the time-dependent dynamics. The key subject of such pump-probe experiments is the visualization and understanding of the time-dependent evolution of the studied quantum system [7]. Mastering the art of tracing the dynamics in a system also opens the door to control quantum processes and even chemical reactions in real time.

In molecular systems, vibrational and rotational wave-packet dynamics have been successfully monitored [8–21], with coherent control of the dynamics being the central objective. A key observable for vibrational molecular wave packets, evolving on anharmonic potential curves or surfaces, is their revival time, which has been detected via pump-probe experiments in the past [22–25]. Strong-field control of a vibrational wave packet has been demonstrated in a pump-control-probe scheme, implementing three precisely timed light pulses [26,27]. Given the multi-pulse complexity of these approaches, one may ask: Is there a direct way to observe and control molecular dynamics, without even introducing a probe step by a laser pulse?

In this letter, we demonstrate molecular wave-packet reshaping in the lightest and most rapidly vibrating neutral molecule, $H_2$, in a pump-control scheme requiring just two laser pulses. We use coherent extreme ultraviolet (XUV) light to launch a vibrational wave packet and subsequently apply a strong few-femtosecond (fs) near-infrared (NIR) control pulse of tunable intensity to alter the wave-packet evolution. Employing the technique of real-time reconstruction of the time-dependent dipole response [28] we extract information about the wave-packet revival, using the coherent dipole emission of the molecule itself as a probe of its time evolution. The strong field is found to influence the revival time, in particular shifting it to earlier times. Even though the NIR field is applied only a few femtoseconds after the initial excitation of the wave packet, the laser pulse influences its dynamical evolution on a much longer time scale.

The experimental setup for transient absorption spectroscopy used to obtain the presented results is described in detail in [29]. In short, few-cycle pulses, ~5 fs full width at half maximum (FWHM) duration, with a central wavelength of 750 nm are focused into xenon gas, where XUV radiation is generated by high-harmonic generation (HHG). An interferometric split-and-delay unit together with concentric filters is used to separate the XUV and NIR pulses both in time and space before they are jointly refocused in the target cell filled with $H_2$ gas at a pressure of 10 mbar over 3 mm interaction length. As depicted in Fig. 1(a), the experiment was conducted in a geometry with a fixed time delay between the XUV and NIR pulse, set to $\tau = 7$ fs, with the XUV pulse arriving first. An additional control parameter for the NIR pulse is its intensity, which can be changed after the HHG and before being focused into the target cell. The maximal intensity of the used NIR pulses is ~$10^{13}$ W/cm$^2$.

After the spectral filtering with a 200-nm-thin indium filter the XUV radiation covers a spectral range between 13 eV and 17 eV of photon energy. In this energy interval, it is possible to reach electronically and vibrationally (i.e. vibronic) excited states of the neutral molecule. In the obtained absorption spectra, we identify resonances of the excited $C\ ^1\Pi_u$, and $D\ ^1\Pi_u$ vibronic bands in neutral H$_2$, the optical density (OD) is shown in Fig. 1(b). After being launched onto different potential energy curves (PECs), the corresponding vibrational wave packets start oscillating. Following the initial dephasing of the wave packet, a revival will occur, as experimentally demonstrated in the past [22–24]. Here, to access the time-dependent dynamics of the excited vibrational wave packets from the measured absorption spectrum, specifically of the excited $D$-state vibrational wave packet, we reconstruct the time-dependent dipole response of the system. For a sufficiently short excitation pulse it was shown in [28] that the coherent dipole emission $d(t)$ can be reconstructed from a single absorption spectrum $A(\omega)$ according to the equation

$$d(t) \propto \mathfrak{F}^{-1}[iA(\omega)](t) = \frac{1}{2\pi}\int_{-\infty}^{+\infty} iA(\omega)e^{-i\omega t}d\omega \quad \text{for } t > 0, \qquad (1)$$

where $\mathfrak{F}^{-1}$ denotes the inverse Fourier transform. By selecting the energy region of only the $D$ band through the orange band-pass filter in Fig. 1(b), we can thus reconstruct the dipole emission, associated to the coherently excited vibrational wave packet of the $D$ state, Fig. 1(c). The reconstructed dipole amplitude exhibits a complicated structure for times up to a few hundred femtoseconds. Owing to the finite spectral resolution of 3 meV at 15 eV photon energy, we can reconstruct the dipole up to a real time of around 400 fs. This is sufficiently long to observe a signature of the first revival of the vibrational wave packet, identified as a periodic peak structure in the dipole amplitude around 270 fs, see inset of Fig.1(c). Individual peaks around the revival time are separated by 18 fs, corresponding to the vibrational period in the $D$ potential energy curve.

To understand how the dipole emission captures the dynamics of the wave packet, we consider a multi-level system and solve the time-dependent Schrödinger equation (TDSE) to model its time evolution. The TDSE is solved by expanding the time-dependent wave function in a basis of field-free Born-Oppenheimer (BO) eigenstates yielding a set of coupled differential equations. The eigenstates included in the model adopted in this work are the following: the absolute ground $X\ ^1\Sigma_g^+$ state with rovibrational quantum numbers $J = 0$ and $v = 0$, the bound states of the electronically excited $D\ ^1\Pi_u$ state, with $J = 1$ and $v = 0, ..., 17$, as well as the bound states with $J = 0$ and $v = 0, ..., 33$ of the electronically excited $EF\ ^1\Sigma_g^+$ potential-energy curve. The eigenenergy of all the included field-free eigenstates in the multi-level model as well as the spatial representation of the bound nuclear wave functions are obtained by solving the time-independent radial Schrödinger equation of nuclear motion using a B-spline basis (600 B-spline functions of order 10 on a linear knot sequence in a radial box with $R_{\max} = 30$ a.u.) and fixed-boundary conditions within the Born-Oppenheimer approximation with the PECs given in [30–32]. The bound nuclear wave functions are calculated for a radial size of $R_{\text{bound}} = 24$ a.u. With the obtained wave functions and the electronic dipole transition moments given in [33,34] the transition dipole moments between all the BO field-free eigenstates contained within the model were calculated and compared for consistency to the values reported in literature [35–38]. Those dipole transition moments, together with the energies, are then used to solve the TDSE describing the molecule exposed to the XUV and NIR light fields, both defined as Gaussian-enveloped pulses with the following parameters: central photon energy $\hbar\omega_{\text{NIR}} = 1.6$ eV and time-domain FWHM$_{\text{NIR}} = 5$ fs for the NIR pulse, and $\hbar\omega_{\text{XUV}} = 14$ eV and time-domain FWHM$_{\text{XUV}} = 0.5$ fs for the XUV pulse, respectively. The laser interaction is considered in the length gauge of the dipole approximation and only dipole-allowed couplings between the states are included. We obtain the solution of the TDSE at each time step, $dt = 0.5$ a.u., via a split-step algorithm method of second-order accuracy. More details on the simulation methods are presented in the supplement material.

All levels included in the model simulation are depicted as horizontal blue lines at their respective energy position in Fig. 2(a). Here, a vertical transition from the ground state (green) to the $D$ potential-energy curve is initiated by the XUV pulse, violet arrow. For an impulsive XUV excitation, within the Franck-Condon (FC) picture, the initially excited superposition of the bound $D$ vibrational states $\Psi_D = \sum_{v=0}^{17} c_v(t_0) \psi_{D,v}$ naturally resembles a copy of the ground state. This intuitive expectation is observed in our simulation, even without explicitly considering the Franck-Condon principle but, on the contrary, fully relying on the transition dipole moments from the ground state to the excited states. To allow for the NIR interaction we include in the model the bound states of the $EF\ ^1\Sigma_g^+$ potential-energy curve, as stated above. As they exhibit the same symmetry as the ground state, they cannot be excited by the XUV pulse, but allow for transitions from and to the $D$ vibronic states via the NIR field.

The field-free time-evolution of the $D$ vibrational wave packet $\Psi_D(t)$ is show in Fig. 2(b). After its initial excitation around internuclear distance $R = 0.76$ Å, a vibrational oscillation in the PEC of the $D$ state evolves. Each of the vibrational coefficients of the $D$-states superposition, in the field-free case: $c_v(t) = e^{-iE_v t/\hbar} c_v(t = t_0)$, develops with its eigenenergy $E_v$. Due to the anharmonicity of the $D$ potential-energy curve, the eigenenergies of the states in the wave packet are not equidistantly spaced, resulting in dephasing and revival. The first revival of the the $D$ vibrational WP occurs at 270 fs. To make a connection to the time-dependent dipole amplitude, which is non-zero only for the overlap of the full (electronic and nuclear) wavefunction with the ground state, we compute the Franck-Condon overlap integral $\int_0^\infty \Psi_{\text{ground}}^*(t)\Psi_D(t)dR$ (dotted line in Fig. 2(c)). The overlap integral and the dipole amplitude between the ground state and the vibrational wave packet match almost perfectly for all times. Furthermore, we compute the integral $\int_{R_1}^{R_2} \Psi_D^*(t)\Psi_D(t)dR$ of the vibrational WP inside the FC region (black line in Fig. 2(c)), with $R_1 = 0.638$ Å and $R_2 = 0.889$ Å. Its comparison to the dipole amplitude further confirms that the peaks of the dipole amplitude coincide with times when the vibrational wave packet localizes inside the Franck-Condon region. It is thus possible to access time information about the internuclear wave packet via the dipole emission. In other words, the molecular ground state acts as an intrinsic probe of the vibrational wave packet and no additional laser pulse is required to probe the time evolution of the system. A second interacting pulse, here an intense NIR pulse, can then be used as a control pulse.

After understanding how the vibrational wave-packet dynamics is encoded in the electronic dipole emission, we now present the experimental observation of impulsive strong-field wave-packet reshaping. Fixing the NIR pulse to the time delay $\tau = 7$ fs after the XUV excitation and gradually tuning its intensity, we record XUV absorption spectra at different NIR intensities in the range between $10^{11}$ and $10^{13}$ W/cm$^2$. In recent experiments, transitions in the energy range of the singly excited vibronic resonances in $H_2$ have been studied for a fixed NIR intensity by scanning its time delay [17,18], observing beatings in absorption spectra and reconstructing vibrational wave packets within a 10-fs-short region of scanned time delay. Here, we do not only have access to the wave-packet dynamics on a 300-fs-long time-scale, up to its first revival, but deliberately change it by a control NIR field introduced at a fixed time after the birth of the wave packet. Figure 3(a) shows the absorption spectra in the energy region of the $D$ vibronic band - the darker the color, the higher the NIR intensity. The lineshapes of the resonances change with increasing NIR intensity. The higher the intensity, the broader the resonances become and their lineshape also assumes a more symmetrical form. Such line-shape changes are known and understood for isolated states in atoms [39,40], but their interpretation for molecular dynamics remained elusive. A reconstruction of the time-dependent dipole emission amplitude centered at the energy of the $D$ vibronic band for different intensities of the NIR control field is shown in Fig. 3(b). Most notably, the structure of the dipole amplitude in the wave-packet revival region between 235 and 300 fs changes. In particular, the peaks associated with the localization of the wave packet in the Franck-Condon region shift to earlier times and become even more prominent for higher intensity. This trend is in qualitative agreement with calculations of the multi-level model simulation, depicted in Fig. 3(c) for increasing NIR intensity. Even though the NIR pulse interacts with the system only for a short time (impulsively) and right after the wave-packet excitation, we find experimental evidence that the subsequent wave-packet evolution is coherently modified for times at least up to its first revival. We thus demonstrate control over the coherent vibrational dynamics of the neutral $H_2$ molecule in an electronically excited state with an approach, in which the time information is revealed in the measurement by the system itself, since the molecular ground state probes the excitation intrinsically.

When an intense NIR field interacts with the molecular system, for the duration of the NIR pulse, the $\psi_{D,v}$ states are strongly coupled to the $\psi_{EF,v}$ states, such that population transfer and field-induced energy shifts occur. The 5-fs-short NIR pulse arrives 7 fs after the XUV excitation-field, Fig. 2(d). The local-in-time distortion of the wave packet through the NIR does have a noticeable effect also for later times. In an impulsive picture, where the pulses correspond to a δ-like action, the change of both amplitude and phase in the $c_v(t)$ coefficients caused by the NIR pulse around $t = \tau$ effectively leads to different initial conditions for their field-free time evolution for all times $t > \tau$. These modified initial conditions therefore impact the complete time evolution of the wave packet. As a result of this control, around the field-free revival time at $t = 270$ fs, we observe the wave packet to reach a local maximum within the FC region approximately 1-2 vibrational periods earlier (i.e. at around 253 fs and 235 fs, respectively), which is illustrated by arrows in Fig. 2(e). Due to the relation between wave-packet and dipole amplitude, the dipole is modified correspondingly, as observed experimentally for different NIR intensities in Fig. 3(b). The impulsive control of the wave packet at a specific time $t = \tau$, with the duration of the interaction being much shorter than the coherent wave-packet evolution, is conceptually similar to the time-domain control of the dipole emission and its imprint in measured absorption spectra for isolated states of atoms [28,40–42]. However, here it collectively affects an entire series of vibrational states reflecting in modified molecular wave-packet dynamics.

In conclusion, we have experimentally measured the reshaping of a molecular wave packet under the influence of a strong and short laser field. Even if interacting with the system only briefly and for a short time after the birth of the wave packet, the NIR field affects its subsequent time evolution reaching up to at least the first revival time. We have demonstrated that by tuning the NIR intensity we can influence when the wave packet rephases. While a wave-packet creation pulse and a control strong-field pulse are required, most notably no probe pulse is needed. We anticipate that additional wavelength control of the intense fields, or even more complex-shaped laser pulses (e.g., through the introduction of chirps) could create almost arbitrary sets of wave-packet coefficients and enhanced controllability of internuclear wave packets up to the fastest (hydrogen) oscillations, also in complex (organic) molecules. Thus, one can use intense laser fields as flexible control knobs to coherently steer the recovery of vibrational wave packets on electronically excited potential energy surfaces to critical points along a reaction coordinate (e.g. conical intersections), and with this initiating a chemical reaction at a desired later time becomes possible.

**Acknowledgements:**

We thank Nikola Mollov for support with the laser setup. We acknowledge support by the Heidelberg STRUCTURES Excellence Cluster funded by the Deutsche Forschungsgemeinschaft (DFG, German Research Foundation) under Germany´s Excellence Strategy EXC 2181/1 - 390900948. P.B.B. acknowledges that her contribution to the project that gave rise to these results received the support of a fellowship from "la Caixa" Foundation (ID 100010434), fellowship code LCF/BQ/EU20/11810064.

**References:**

[1]    E. Goulielmakis et al., *Real-Time Observation of Valence Electron Motion*, Nature **466**, 739 (2010).

[2]   J. Mauritsson et al., *Attosecond Electron Spectroscopy Using a Novel Interferometric Pump-Probe Technique*, Phys Rev Lett **105**, 053001 (2010).

[3]   M. Holler, F. Schapper, L. Gallmann, and U. Keller, *Attosecond Electron Wave-Packet Interference Observed by Transient Absorption*, Phys Rev Lett **106**, 123601 (2011).

[4]   C. Ott et al., *Reconstruction and Control of a Time-Dependent Two-Electron Wave Packet*, Nature **516**, 374 (2014).

[5]   G. Sansone et al., *Electron Localization Following Attosecond Molecular Photoionization*, Nature **465**, 763 (2010).

[6]   F. Calegari et al., *Ultrafast Electron Dynamics in Phenylalanine Initiated by Attosecond Pulses*, Science (1979) **346**, 336 (2014).

[7]   A. H. Zewail, *Femtochemistry: Atomic-Scale Dynamics of the Chemical Bond*, Journal of Physical Chemistry A **104**, 5660 (2000).

[8]   M. Dantus, M. H. M. Janssen, and A. H. Zewail, *Femtosecond Probing of Molecular Dynamics by Mass-Spectrometry in a Molecular Beam*, Chem Phys Lett **181**, 281 (1991).

[9]   T. Baumert, M. Grosser, R. Thalweiser, and G. Gerber, *Femtosecond Time-Resolved Molecular Multiphoton Ionization: The $Na_2$ System*, Phys Rev Lett **67**, 3753 (1991).

[10]  B. J. Sussman, D. Townsend, M. Yu. Ivanov, and A. Stolow, *Dynamic Stark Control of Photochemical Processes*, Science (1979) **314**, 278 (2006).

[11]  B. Feuerstein, T. Ergler, A. Rudenko, K. Zrost, C. D. Schröter, R. Moshammer, J. Ullrich, T. Niederhausen, and U. Thumm, *Complete Characterization of Molecular Dynamics in Ultrashort Laser Fields*, Phys Rev Lett **99**, 153002 (2007).

[12]  F. Kelkensberg et al., *Molecular Dissociative Ionization and Wave-Packet Dynamics Studied Using Two-Color XUV and IR Pump-Probe Spectroscopy*, Phys Rev Lett **103**, 16 (2009).

[13]  D. Brinks, F. D. Stefani, F. Kulzer, R. Hildner, T. H. Taminiau, Y. Avlasevich, K. Müllen, and N. F. van Hulst, *Visualizing and Controlling Vibrational Wave Packets of Single Molecules*, Nature **465**, 905 (2010).

[14]  S. De et al., *Following Dynamic Nuclear Wave Packets in $N_2$, $O_2$, and CO with Few-Cycle Infrared Pulses*, Phys Rev A **84**, 1 (2011).

[15]  T. Okino, Y. Furukawa, Y. Nabekawa, S. Miyabe, A. Amani Eilanlou, E. J. Takahashi, K. Yamanouchi, and K. Midorikawa, *Direct Observation of an Attosecond Electron Wave Packet in a Nitrogen Molecule*, Sci Adv **1**, (2015).

[16]  E. R. Warrick, W. Cao, D. M. Neumark, and S. R. Leone, *Probing the Dynamics of Rydberg and Valence States of Molecular Nitrogen with Attosecond Transient Absorption Spectroscopy*, Journal of Physical Chemistry A **120**, 3165 (2016).

[17]  Y. Cheng, M. Chini, X. Wang, A. González-Castrillo, A. Palacios, L. Argenti, F. Martín, and Z. Chang, *Reconstruction of an Excited-State Molecular Wave Packet with Attosecond Transient Absorption Spectroscopy*, Phys Rev A (Coll Park) **94**, 023403 (2016).

[18] W. Cao, E. R. Warrick, A. Fidler, S. R. Leone, and D. M. Neumark, *Excited-State Vibronic Wave-Packet Dynamics in $H_2$ Probed by XUV Transient Four-Wave Mixing*, Phys Rev A (Coll Park) **97**, 1 (2018).

[19] H. Timmers, X. Zhu, Z. Li, Y. Kobayashi, M. Sabbar, M. Hollstein, M. Reduzzi, T. J. Martínez, D. M. Neumark, and S. R. Leone, *Disentangling Conical Intersection and Coherent Molecular Dynamics in Methyl Bromide with Attosecond Transient Absorption Spectroscopy*, Nat Commun **10**, 3133 (2019).

[20] E. T. Karamatskos et al., *Molecular Movie of Ultrafast Coherent Rotational Dynamics of OCS*, Nat Commun **10**, 3364 (2019).

[21] P. Peng, Y. Mi, M. Lytova, M. Britton, X. Ding, A. Y. Naumov, P. B. Corkum, and D. M. Villeneuve, *Coherent Control of Ultrafast Extreme Ultraviolet Transient Absorption*, Nat Photonics **16**, 45 (2022).

[22] M. J. J. Vrakking, D. M. Villeneuve, and A. Stolow, *Observation of Fractional Revivals of a Molecular Wave Packet*, Phys Rev A (Coll Park) **54**, R37 (1996).

[23] Th. Ergler, A. Rudenko, B. Feuerstein, K. Zrost, C. D. Schröter, R. Moshammer, and J. Ullrich, *Spatiotemporal Imaging of Ultrafast Molecular Motion: Collapse and Revival of the $D_2^+$ Nuclear Wave Packet*, Phys Rev Lett **97**, 193001 (2006).

[24] B. Feuerstein, Th. Ergler, A. Rudenko, K. Zrost, C. D. Schröter, R. Moshammer, J. Ullrich, T. Niederhausen, and U. Thumm, *Complete Characterization of Molecular Dynamics in Ultrashort Laser Fields*, Phys Rev Lett **99**, 153002 (2007).

[25] M. Magrakvelidze et al., *Tracing Nuclear-Wave-Packet Dynamics in Singly and Doubly Charged States of $N_2$ and $O_2$ with XUV-Pump-XUV-Probe Experiments*, Phys Rev A **86**, 1 (2012).

[26] W. A. Bryan et al., *Redistribution of Vibrational Population in a Molecular Ion with Nonresonant Strong-Field Laser Pulses*, Phys Rev A (Coll Park) **83**, 021406 (2011).

[27] E. R. Warrick, A. P. Fidler, W. Cao, E. Bloch, D. M. Neumark, and S. R. Leone, *Multiple Pulse Coherent Dynamics and Wave Packet Control of the $N_2$ A'' $^1\Sigma^+_g$ Dark State by Attosecond Four-Wave Mixing*, Faraday Discuss **212**, 157 (2018).

[28] V. Stooß, S. M. Cavaletto, S. Donsa, A. Blättermann, P. Birk, C. H. Keitel, I. Březinová, J. Burgdörfer, C. Ott, and T. Pfeifer, *Real-Time Reconstruction of the Strong-Field-Driven Dipole Response*, Phys Rev Lett **121**, 173005 (2018).

[29] V. Stooß, M. Hartmann, P. Birk, G. D. Borisova, T. Ding, A. Blättermann, C. Ott, and T. Pfeifer, *XUV-Beamline for Attosecond Transient Absorption Measurements Featuring a Broadband Common Beam-Path Time-Delay Unit and in Situ Reference Spectrometer for High Stability and Sensitivity*, Review of Scientific Instruments **90**, 053108 (2019).

[30] T. E. Sharp, *Potential-Energy Curves for Molecular Hydrogen and Its Ions*, At Data Nucl Data Tables **2**, 119 (1970).

[31] L. Wolniewicz and K. Dressler, *The B $^1\Sigma^+_u$, B' $^1\Sigma^+_u$, C $^1\Pi_u$, and D $^1\Pi_u$ States of the H₂ Molecule. Matrix Elements of Angular and Radial Nonadiabatic Coupling and Imp*, J Chem Phys **88**, 3861 (1988).

[32] H. Nakashima and H. Nakatsuji, *Solving the Schrödinger Equation of Hydrogen Molecule with the Free Complement-Local Schrödinger Equation Method: Potential Energy Curves of the Ground and Singly Excited Singlet and Triplet States, Σ, Π, Δ, and Φ*, Journal of Chemical Physics **149**, (2018).

[33] L. Wolniewicz and G. Staszewska, *$^1\Sigma^+_u \rightarrow X^1\Sigma^+_g$ Transition Moments for the Hydrogen Molecule*, J Mol Spectrosc **217**, 181 (2003).

[34] L. Wolniewicz and G. Staszewska, *Excited $^1\Pi_u$ States and the $^1\Pi_u \rightarrow X^1\Sigma^+_g$ Transition Moments of the Hydrogen Molecule*, J Mol Spectrosc **220**, 45 (2003).

[35] M. Glass-Maujean, P. Quadrelli, K. Dressler, and L. Wolniewicz, *Transition Probabilities for Spontaneous Emission in the Adiabatic and Nonadiabatic Approximations for All Bound Vibrational Levels of the E,F$^1\Sigma^+_g$, G,K$^1\Sigma^+_g$, and H,H"$^1\Sigma_g$*, Phys Rev A (Coll Park) **28**, 2868 (1983).

[36] M. Glass-Maujean, *Transition Probabilities for the D and B' Vibrational Levels to the X Vibrational Levels and Continuum of H₂*, At Data Nucl Data Tables **30**, 301 (1984).

[37] M. Glass-Maujean, P. Quadrelli, and K. Dressler, *Band Transition Moments between Excited Singlet States of the H₂ Molecule, Nonadiabatic Eigenvectors, and Probabilities for Spontaneous Emission*, At Data Nucl Data Tables **30**, 273 (1984).

[38] W. F. Chan, G. Cooper, and C. E. Brion, *Absolute Optical Oscillator Strengths (11-20 EV) and Transition Moments for the Photoabsorption of Molecular Hydrogen in the Lyman and Werner Bands*, Chem Phys **168**, 375 (1992).

[39] C. Ott, A. Kaldun, P. Raith, K. Meyer, M. Laux, J. Evers, C. H. Keitel, C. H. Greene, and T. Pfeifer, *Lorentz Meets Fano in Spectral Line Shapes: A Universal Phase and Its Laser Control*, Science (1979) **340**, 716 (2013).

[40] A. Kaldun, C. Ott, A. Blättermann, M. Laux, K. Meyer, T. Ding, A. Fischer, and T. Pfeifer, *Extracting Phase and Amplitude Modifications of Laser-Coupled Fano Resonances*, Phys Rev Lett **112**, 103001 (2014).

[41] C. Ott, A. Kaldun, P. Raith, K. Meyer, M. Laux, J. Evers, C. H. Keitel, C. H. Greene, and T. Pfeifer, *Lorentz Meets Fano in Spectral Line Shapes: A Universal Phase and Its Laser Control*, Science (1979) **340**, 716 (2013).

[42] A. Blättermann, C. Ott, A. Kaldun, T. Ding, V. Stooß, M. Laux, M. Rebholz, and T. Pfeifer, *In Situ Characterization of Few-Cycle Laser Pulses in Transient Absorption Spectroscopy*, Opt Lett **40**, 3464 (2015).

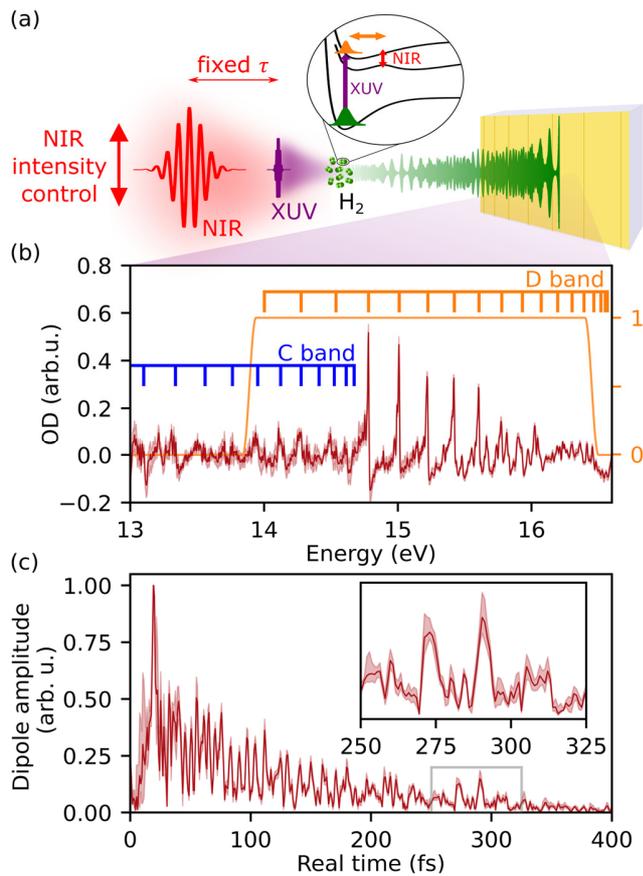

Fig. 1: Reconstruction of the time-dependent dipole response of a vibrational wave packet in molecular hydrogen. (a) Experimental scheme for the excitation of a vibrational WP in H$_2$ via an XUV pulse (violet) and its subsequent coupling with an NIR pulse (red) of controllable intensity. (b) Absorption data (OD) for negligible NIR intensity with marked positions of the $C$- and $D$-band vibronic resonances, blue and orange respectively. (c) Reconstructed time-dependent dipole emission amplitude from the spectral data inside the reconstruction window including only the $D$ vibrational band (shown in orange in (b)). The inset for real time between 250 fs and 325 fs shows the dipole amplitude around the revival time of the $D$ vibrational WP.

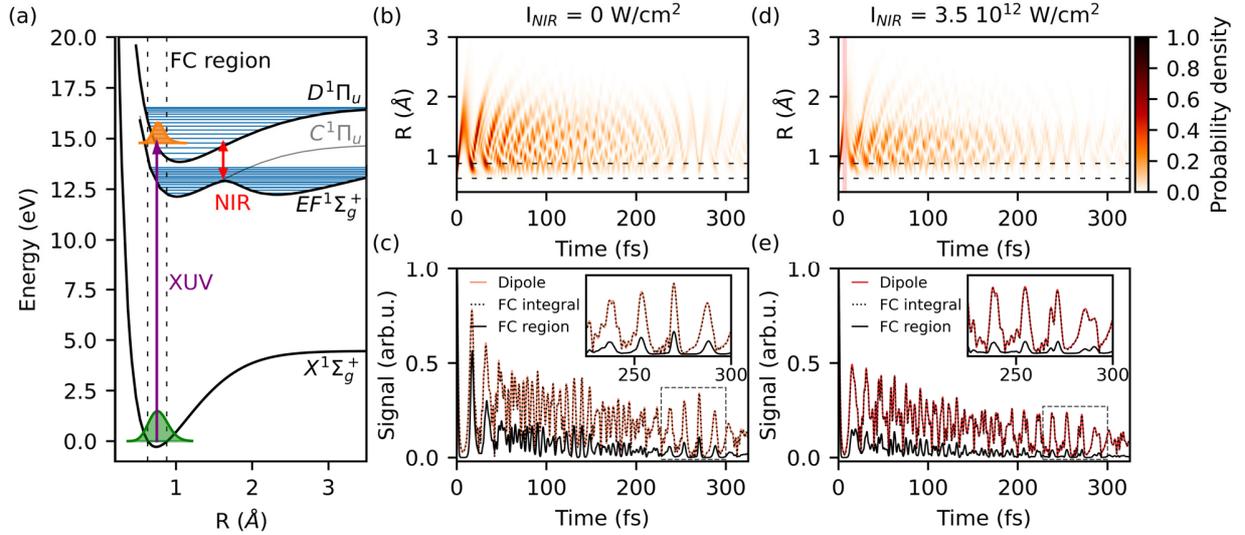

Fig. 2: Multi-level simulation of the vibrational dynamics in molecular hydrogen influenced by a strong NIR field. (a) Relevant potential-energy curves of the $H_2$ molecule and their corresponding bound states (horizontal blue lines) included in the model simulation. The XUV pulse excites a vibronic wave packet in the $D\ ^1\Pi_u$ state, which can couple via an NIR pulse to the $EF\ ^1\Sigma_g^+$ vibrational states. (b) Calculated time-evolution of the excited $D$-band wave packet, obtained for no NIR intensity and (d) for $I_{NIR} = 3.5\ 10^{12}$ W/cm$^2$. (c) and (e) Time-dependent dipole amplitude, Franck-Condon overlap integral and integral inside the FC region between 0.638 Å and 0.889 Å without and with NIR field, respectively. The NIR field couples the $D$ and $EF$ vibrational states shortly (7 fs, light-red region in (d)) after the XUV wave-packet excitation, and changes the WP revival (around 270 fs), observed in both the dipole amplitude and the FC integral.

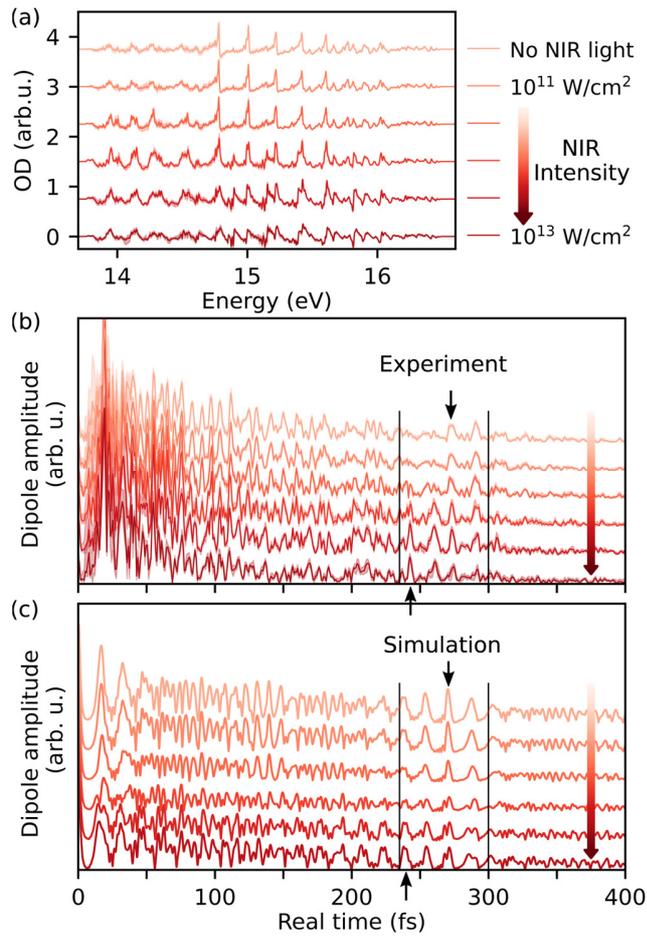

Fig. 3: NIR-intensity-induced change of the vibrational wave-packet dynamics. (a) Measured absorption spectrum in the region of the *D* band for NIR intensities between $10^{11}$ (top) and $10^{13}$ W/cm² (bottom), shown after applying the reconstruction window from Fig. 1(b). (b) and (c) Reconstructed and calculated time-dependent dipole emission amplitude of the *D* vibrational wave for different NIR intensities, respectively. The vertical black lines show the region around the revival of the wave packet. Black arrows point to the dipole maxima associated to the revival of the wave packet, shifting to earlier times with increasing NIR intensity. All subfigures share the same intensity colorscale. For better visibility the individual lines are ordered from lower ($10^{11}$ W/cm²) intensity at the top to higher ($10^{13}$ W/cm²) intensity at the bottom for all presented data.

# Laser control of an excited-state vibrational wave packet in neutral H$_2$

# Supplemental information

Gergana D. Borisova[1]*, Paula Barber Belda[1], Shuyuan Hu[1], Paul Birk[1], Veit Stooß[1],
Maximilian Hartmann[1], Daniel Fan[1], Robert Moshammer[1], Alejandro Saenz[2],
Christian Ott[1]° and Thomas Pfeifer[1]#

*borisova@mpi-hd.mpg.de, °christian.ott@mpi-hd.mpg.de, #thomas.pfeifer@mpi-hd.mpg.de

[1]Max-Planck-Institut für Kernphysik, Saupfercheckweg 1, 69117 Heidelberg, Germany
[2]Institut für Physik, Humboldt-Universität zu Berlin, 12489 Berlin, Germany

# 1 Details of the data evalutaion

## 1.1 Data averaging and error estimation

The presented data in the main text is a result of five individual measurements for each of the six considered NIR intensities around the time-delay $\tau = 7$ fs between the XUV and the NIR pulse. The considered time-delay region around the chosen fixed time delay is a quarter of an NIR-pulse cycle, i.e. a data set at five different time steps of 0.17 fs around the chosen time-delay position is built. The absorption OD data is then a result of the averaging of all $5 \times 5$ measurements for each NIR intensity. The estimation of the error bars for the absorption OD data is determined as the statistical error of the mean over all individual experiments at each NIR intensity. The shown reconstructed dipole is again the mean over all reconstructions of all individual $5 \times 5$ experiments. The error bars of the reconstruction are estimated with an upper and lower value, respectively, the highest and lowest reconstructed value of the dipole amplitude at a given real time among all dipole reconstructions used to calculate the mean at each NIR intensity. This calculation of the error of the dipole amplitude is more conservative than simply taking the statistical error estimation. All error bars are shown in the pictures of the main text as a shading around the plotted curves.

## 1.2 Reconstruction window

The used reconstruction window, shown in Fig. 1(b) of the main text, is a band-pass filter with $\cos^2$-shaped rising and falling edge over an energy region of 0.1 eV and is otherwise constant 1 between the two edges and vanishes outside them. The reconstruction window is applied multiplicatively to the measured absorption data.

## 1.3 NIR Intensity calibration

The NIR intensity calibration is performed through a measurement of the beam profile and pulse energy in the interaction focus, and a dispersion scan (D-Scan) measurement of the pulse duration, leading to the intensity calibration shown in Fig. S1. The error of the estimated NIR intensity is an order of magnitude smaller than each value, which is why on the logarithmic scale the error bars appear smaller than the data points.

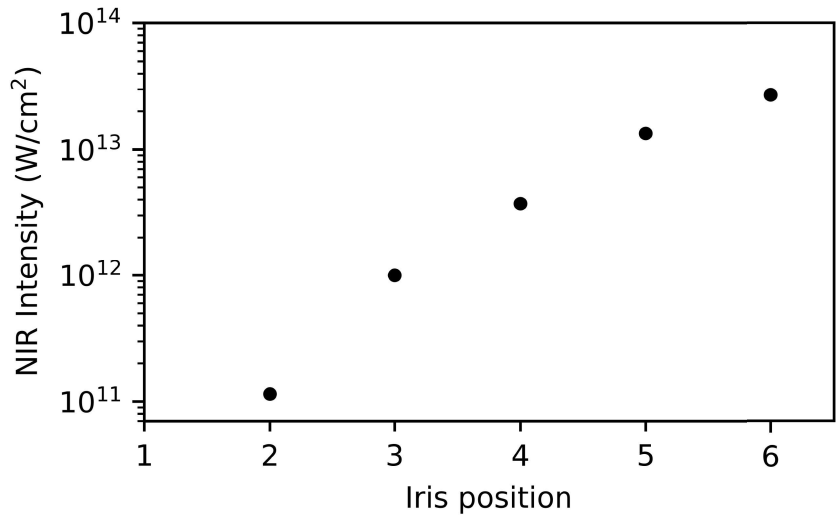

Figure 1: Intensity calibration, with iris position 1 corresponding to no NIR intensity.

# 2 Simulation details

## 2.1 Calculation of the dipole transition elements

Throughout this section, atomic units (a.u.) are used, unless stated explicitly otherwise. The simulation adopts the Born-Oppenheimer approximation, in which the total wave function of the model system reads

$$\Psi(\boldsymbol{r}_1, \boldsymbol{r}_2; \boldsymbol{R}) = \phi(\boldsymbol{r}_1, \boldsymbol{r}_2; R)\psi(\boldsymbol{R}) \equiv \phi(\boldsymbol{r}; R)\psi(\boldsymbol{R}).$$

The solutions of the radial Schrödinger equation are the nuclear wave functions for a given potential-energy curve, in our case, those of the molecular ground state and of the $D$ and $EF$ electronic excited states. We then calculate the transition probabilities between the energy levels for all considered dipole transitions. If $\hat{W}$ is the transition operator, a transition matrix element from an initial to a final state is given by

$$T_{i \to f} = \langle \Psi_f(\boldsymbol{r}, \boldsymbol{R}) | \hat{W} | \Psi_i(\boldsymbol{r}, \boldsymbol{R}) \rangle \approx \langle \Psi_f(\boldsymbol{r}, \boldsymbol{R}) | \hat{d} | \Psi_i(\boldsymbol{r}, \boldsymbol{R}) \rangle \equiv d_{i \to f}, \quad (1)$$

with $\hat{d}$ being the dipole operator. The square of the matrix element gives the transition probability between two states:

$$P_{i \to f} = |T_{i \to f}|^2 \approx \left| \langle \Psi_f(\boldsymbol{r}, \boldsymbol{R}) | \hat{d} | \Psi_i(\boldsymbol{r}, \boldsymbol{R}) \rangle \right|^2. \tag{2}$$

Recalling the BO-separation of the nuclear and electronic wave functions, the transition elements become

$$d_{i \to f} = \langle \psi_f(\boldsymbol{R}) | \langle \phi_f(\boldsymbol{r}; R) | \hat{d} | \phi_i(\boldsymbol{r}; R) \rangle | \psi_i(\boldsymbol{R}) \rangle, \tag{3}$$

with $\langle \phi_f(\boldsymbol{r}; R) | \hat{d} | \phi_i(\boldsymbol{r}; R) \rangle \equiv C_{if}(R)$ called the electronic coupling matrix elements, or short the *couplings*. In the Franck-Condon picture the couplings are independent of the internuclear separation $R$, leading to

$$d_{i \to f} = \langle \phi_f(\boldsymbol{r}; R) | \hat{d} | \phi_i(\boldsymbol{r}; R) \rangle \langle \psi_f(\boldsymbol{R}) | \psi_i(\boldsymbol{R}) \rangle = \tilde{C}_{if} \langle \psi_f(\boldsymbol{R}) | \psi_i(\boldsymbol{R}) \rangle \tag{4}$$

as the transition elements, i.e. only considering the overlap between the nuclear wave functions of the involved levels, which are called the Franck-Condon factors. This approximation establishes the applicability of the dipole reconstruction method applied to the measured absorption data. In the model simulation, however, we include the dependency of the couplings on $R$ found in literature [1,2], arriving at

$$d_{i \to f} = \langle \psi_f(\boldsymbol{R}) | C_{if}(R) | \psi_i(\boldsymbol{R}) \rangle. \tag{5}$$

## 2.2 Hamiltonian of the multi-level model simulation

The Hamiltonian of the multi-level system is the sum of the field-free Hamiltonian $\mathcal{H}_0$ and the time-dependent interaction Hamiltonian, $\mathcal{H}_{int}(t)$:

$$\mathcal{H}(t) = \underbrace{\mathcal{H}_0}_{\substack{\text{eigenenergies} \\ \text{(diag.)}}} + \underbrace{\mathcal{H}_{\text{int}}(t)}_{\substack{\text{laser interaction} \\ \text{(not diag.)}}} \tag{6}$$

In the eigenstate basis, $\mathcal{H}_0$ is a diagonal matrix, with its elements being the energy levels of the different electronic vibrational levels $E_{\Theta_\nu}$, where $\Theta$ is the electronic state and $\nu$ the vibrational quantum number.

The interaction Hamiltonian $\mathcal{H}_{\text{int}}(t)$ includes the interactions with both the XUV and the NIR pulses:

$$\mathcal{H}_{\text{int}} = \mathcal{H}_{\text{int,XUV}} + \mathcal{H}_{\text{int,NIR}}. \tag{7}$$

It thus introduces the non-diagonal terms responsible for the transitions between the levels. In the dipole approximation, the light-matter interaction in $\mathcal{H}_{\text{int}}$ is given by

$$\mathcal{H}_{\text{int},ij} = d_{i \to j} \mathcal{E}(t). \tag{8}$$

In our model, the transitions from the ground state are introduced by the XUV field, i.e. mathematically include $\mathcal{E}_{XUV}(t) \equiv \mathcal{E}(t)$, and the $D$ and $EF$ states couple through the NIR electric field, $\mathcal{E}_{NIR}(t) \equiv \mathcal{E}(t)$. With the following color code adopted: blue for the diagonal elements of the free Hamiltonian, violet for the XUV transitions included in $H_{\text{int},XUV}$ and red for the NIR-mediated transitions, which come from $H_{\text{int, NIR}}$, the total Hamiltonian reads

$$\mathcal{H}(t) = \mathcal{H}_0 + \mathcal{H}_{\text{int,XUV}} + \mathcal{H}_{\text{int,NIR}}, \tag{9}$$

which in its explicit form corresponds to

$$\mathcal{H}(t) = \begin{pmatrix} E_{X_0} & d_{X \leftrightarrow D_0}\mathcal{E}(t) & d_{X \leftrightarrow D_1}\mathcal{E}(t) & \cdots & 0 & 0 & \cdots & \cdots & 0 \\ d_{X \leftrightarrow D_0}\mathcal{E}(t) & E_{D_0} & 0 & \cdots & d_{D_0 \leftrightarrow EF_0}\mathcal{E}(t) & d_{D_0 \leftrightarrow EF_1}\mathcal{E}(t) & \cdots & \cdots & d_{D_0 \leftrightarrow EF_{33}}\mathcal{E}(t) \\ d_{X \leftrightarrow D_1}\mathcal{E}(t) & 0 & E_{D_1} & \cdots & d_{D_1 \leftrightarrow EF_0}\mathcal{E}(t) & d_{D_1 \leftrightarrow EF_1}\mathcal{E}(t) & \cdots & \cdots & d_{D_1 \leftrightarrow EF_{33}}\mathcal{E}(t) \\ \vdots & \vdots & \vdots & \ddots & \vdots & \vdots & \vdots & \vdots & \vdots \\ 0 & d_{D_0 \leftrightarrow EF_0}\mathcal{E}(t) & d_{D_1 \leftrightarrow EF_0}\mathcal{E}(t) & \cdots & E_{EF_0} & 0 & \cdots & \cdots & 0 \\ 0 & d_{D_0 \leftrightarrow EF_1}\mathcal{E}(t) & d_{D_1 \leftrightarrow EF_1}\mathcal{E}(t) & \cdots & 0 & E_{EF_1} & 0 & \cdots & 0 \\ \vdots & \vdots & \vdots & \vdots & \vdots & \vdots & \ddots & \vdots & \vdots \\ 0 & d_{D_0 \leftrightarrow EF_{33}}\mathcal{E}(t) & d_{D_1 \leftrightarrow EF_{33}}\mathcal{E}(t) & \cdots & 0 & 0 & 0 & \cdots & E_{EF_{33}} \end{pmatrix}, \tag{10}$$

where the subindex of the states refers to the vibrational quantum number $\nu$.

To account for the finite life time of the excited states, we include an imaginary term in the free Hamiltonian energies, which introduces an exponential decay of the population of the states. For a lifetime of $\frac{1}{\Gamma}$, the corresponding transformation is

$$E \leftarrow E - i\frac{\Gamma}{2}. \tag{11}$$

The ground state $X_{\nu=0}$ has no decay rate.

### 2.3 Solving the time-dependent Schrödinger Equation

In the model simulation, we solve numerically the time-dependent Schrödinger Equation for the chose states, as described in the main text:

$$\mathcal{H}(t) \left|\Psi(t)\right\rangle = i\frac{\partial}{\partial t} \left|\Psi(t)\right\rangle. \tag{12}$$

Numerically, the state vector $|\Psi(t)\rangle$ is represented in the free basis $\{|i\rangle\}_{i=1}^{N}$, where $N$ is the total number of considered states, which are all eigenvectors of the free Hamiltonian $\mathcal{H}_0$, i.e.,

$$\mathcal{H}_0 |i\rangle = E_{0,i} |i\rangle. \tag{13}$$

where $E_{0,i}$ are the eigenenergies. The expansion of the state vector with the coefficients $c_i(t)$, their squared value being the probability of the system to populate a given state $i$, is as follows:

$$|\Psi(t)\rangle = \sum_{i=0}^{N} c_i(t) |i\rangle. \tag{14}$$

The basis states $\{|i\rangle\}_{i=1}^{N}$ are the solutions of the radial field-free Schrödinger equation, as discussed above. The state vector $\vec{c}(t)$ is then the one to be calculated at each time step,

$$\vec{c}(t) = \begin{pmatrix} c_0(t) \\ c_1(t) \\ \vdots \\ c_N(t) \end{pmatrix}. \tag{15}$$

The solution to the TDSE (12) is an exponential function, in atomic units:

$$\vec{c}(t) = e^{-i\mathcal{H}(t) \cdot t} \vec{c}(0), \tag{16}$$

where $\vec{c}(0)$ represents the initial state vector at $t=0$. In the simulation, the system always starts with the total population in the ground state.

The solution of the TDSE is then the evolution of the system with its time-propagation operator $e^{-i\mathcal{H}t}$ for a total time $T$ divided in steps $\Delta t$. The time-propagation operator is approximated by a second-order split-step-algorithm [3]:

$$e^{-i\mathcal{H}(t) \cdot t} = e^{-i(\mathcal{H}_0 + \mathcal{H}_{\text{int}}(t)) \cdot t} \approx e^{-i\mathcal{H}_0 \cdot \frac{t}{2}} e^{-i\mathcal{H}_{\text{int}}(t) \cdot t} e^{-i\mathcal{H}_0 \cdot \frac{t}{2}}. \tag{17}$$

The basis elements are eigenfunctions of $\mathcal{H}_0$, turning the exponential part $e^{-i\mathcal{H}_0 \cdot \frac{t}{2}}$ into a diagonal matrix, such that one can directly apply

$$|\Psi(t_n)\rangle = e^{-i\mathcal{H}_0 \cdot \frac{\Delta t}{2}} e^{-i\mathcal{H}_{\text{int}}(t_n) \cdot \Delta t} e^{-i\mathcal{H}_0 \cdot \frac{\Delta t}{2}} |\Psi(t_{n-1})\rangle, \tag{18}$$

where the application of the half-time evolution of the free Hamiltonian to the state vector $e^{-i\mathcal{H}_0 \cdot \frac{\Delta t}{2}} |\Psi(t_{n-1})\rangle$ is equivalent to the multiplication of each of the coefficients by its free-energy evolution $e^{-iE_{0,i} \cdot \frac{\Delta t}{2}} c_i(t_{n-1})$ and $\mathcal{H}_{\text{int}}(t_n)$ is the interaction Hamiltonian at time $t_n =$ $t_0 + (n-1)\Delta t$, with $n$ an integer number.

To compute the interaction part, we introduce a separation of the XUV- and NIR-components, accurate up to first order:

$$e^{-i\mathcal{H}_{\text{int}}(t_n)\cdot t} = e^{-i\left(\mathcal{H}_{\text{int,XUV}}(t) + \mathcal{H}_{\text{int,NIR}}(t)\right)\cdot t} \approx e^{-i\mathcal{H}_{\text{int,XUV}}(t)\cdot t}\, e^{-i\mathcal{H}_{\text{int,NIR}}(t)\cdot t}. \tag{19}$$

This enables the independent propagation of both parts of the interaction Hamiltonian, requiring just a single diagonalisation at the beginning of the calculation. The electric field can then be factored out of each of the matrices, and the unitary eigenvectors of $\mathcal{H}_{\text{int,XUV}}$ and $\mathcal{H}_{\text{int,NIR}}$ remain constant in time. Note that the eigenvalues do depend on the electric field, but just as a multiplicative factor. The basis change is then applied

$$\mathcal{H}^D_{\text{int,XUV}}(t) = T^{-1}_{\text{XUV}} \mathcal{H}_{\text{int,XUV}}(t) T_{\text{XUV}}, \tag{20}$$

$$\mathcal{H}^D_{\text{int,NIR}}(t) = T^{-1}_{\text{NIR}} \mathcal{H}_{\text{int,NIR}}(t) T_{\text{NIR}}, \tag{21}$$

where the matrices $T_{\text{XUV}}$ and $T_{\text{NIR}}$ are time independent.

Applying now the time evolution induced from the interaction Hamiltonian and decomposed as in equation (19), the basis change is first applied to transform the state vector in the $\mathcal{H}_{\text{int,XUV}}$-basis

$$\left|\Psi^D_{XUV}(t)\right\rangle = T^{-1}_{\text{XUV}} \left|\Psi(t)\right\rangle. \tag{22}$$

Next, the time evolution is carried out, with the now diagonal matrix, making the operation trivial:

$$\left|\Psi^D_{XUV}(t_n)\right\rangle \leftarrow e^{-i\mathcal{H}^D_{\text{int,XUV}}(t_n)\cdot \Delta t} \left|\Psi^D_{XUV}(t_{n-1})\right\rangle. \tag{23}$$

As a last step, one must return to the original free-Hamiltonian basis, which is achieved by the transformation

$$\left|\Psi(t)\right\rangle = T_{\text{XUV}} \left|\Psi^D_{XUV}(t)\right\rangle. \tag{24}$$

The whole procedure is repeated, now with a basis change to the $\mathcal{H}_{\text{int,NIR}}$ eigenbasis, i.e.,

$$\left|\Psi^D_{NIR}(t)\right\rangle = T^{-1}_{\text{NIR}} \left|\Psi(t)\right\rangle. \tag{25}$$

The time evolution here has the form

$$\left|\Psi^D_{NIR}(t_n)\right\rangle \leftarrow e^{-i\mathcal{H}^D_{\text{int,NIR}}(t_n)\cdot \Delta t} \left|\Psi^D_{NIR}(t_{n-1})\right\rangle. \tag{26}$$

Again, finally one returns to the original free-Hamiltonian basis by transforming back

$$|\Psi(t)\rangle = T_{\text{NIR}} \left|\Psi^D_{NIR}(t)\right\rangle. \tag{27}$$

Following equation (17), the second half of the free-Hamiltonian time evolution must be applied, that means, the state vector must once more be multiplied with the exponential of $\mathcal{H}_0$

$$|\Psi(t_n)\rangle \leftarrow e^{-i\mathcal{H}_0 \cdot \frac{\Delta t}{2}} |\Psi(t_{n-1})\rangle. \tag{28}$$

In summary, the whole time evolution for each time step is given by

$$|\Psi(t_n)\rangle = e^{-i\mathcal{H}_0 \cdot \frac{\Delta t}{2}} T_{\text{NIR}}\, e^{-i\mathcal{H}^D_{\text{int,NIR}}(t_n)\cdot \Delta t}\, T^{-1}_{\text{NIR}}\, T_{\text{XUV}}\, e^{-i\mathcal{H}^D_{\text{int,XUV}}(t_n)\cdot \Delta t}\, T^{-1}_{\text{XUV}}\, e^{-i\mathcal{H}_0 \cdot \frac{\Delta t}{2}} |\Psi(t_{n-1})\rangle. \tag{29}$$

In addition to the state vector, the time-dependent dipole moment (TDDM) was computed at every time step. With respect to the ground state, it is defined as

$$d(t) = \langle\Psi(t)|\hat{\mathbf{d}}|\Psi(t)\rangle = \sum_{i=1}^{N} c_0^*(t) c_i(t) \langle 0 | \hat{\mathbf{d}} | i \rangle + c.c. \tag{30}$$

From the TDDM, the optical density can also be calculated.

## References

[1] L. Wolniewicz and G. Staszewska, $^1\Sigma_u^+ \longrightarrow X^1\Sigma_g^+$ *Transition Moments for the Hydrogen Molecule*, J Mol Spectrosc 217, 181 (2003)
[2] L. Wolniewicz and G. Staszewska, *Excited $^1\Pi_u$ States and the $^1\Pi_u \longrightarrow X^1\Sigma_g^+$ Transition Moments of the Hydrogen Molecule*, J Mol Spectrosc 220, 45 (2003)
[3] A. D. Bandrauk and H. Shen, *Improved exponential split operator method for solving the time-dependent Schrödinger equation*, Chemical Physics Letters 176, 428–432 (1991)